\documentclass[]{spie}  

\usepackage[]{graphicx}
\usepackage{xspace}
\usepackage[]{amsmath}
\usepackage{amssymb}
\usepackage{tablefootnote}

\newcommand{\eg}{{\it e.g.}\xspace}

\newcommand{\etal}{{\it et al.}\xspace}

\title{ATLAST detector needs for direct spectroscopic biosignature characterization in the visible and near-IR} 

\author{
Bernard J. Rauscher,\supit{a}
Matthew R. Bolcar,\supit{a}
Mark Clampin,\supit{a}\\
Shawn D. Domagal-Goldman,\supit{a}
Michael W. McElwain,\supit{a}
S. H. Moseley,\supit{a}
Carl Stahle,\supit{a}\\
Christopher C. Stark,\supit{b} and
Harley A. Thronson\supit{a}
\skiplinehalf
\supit{a}NASA Goddard Space Flight Center, 8800 Greenbelt Road, Greenbelt, MD, 20771, U.S.A.\\
\supit{b}Space Telescope Science Institute, 3700 San Martin Drive, Baltimore, MD, 21218, U.S.A.}

\authorinfo{
Presented 9 August 2015 at SPIE Optics+Photonics, San Diego, CA, U.S.A., Paper 9602-12\\
Further author information: (Send correspondence to B.J.R.)\\B.J.R.: E-mail: Bernard.J.Rauscher@nasa.gov, Telephone: +1 301-286-4871}

\begin{document} 
\maketitle 

\begin{abstract}
Are we alone? Answering this ageless question will be a major focus for astrophysics in coming decades. Our tools will include unprecedentedly large UV-Optical-IR space telescopes working with advanced coronagraphs and starshades. Yet, these facilities will not live up to their full potential without better detectors than we have today. To inform detector development, this paper provides an overview of visible and near-IR (VISIR; $\lambda=0.4-1.8~\mu\textrm{m}$) detector needs for the Advanced Technology Large Aperture Space Telescope (ATLAST), specifically for spectroscopic characterization of atmospheric biosignature gasses. We also provide a brief status update on some promising detector technologies for meeting these needs in the context of a passively cooled ATLAST.
\end{abstract}


\section{INTRODUCTION}\label{Sec:Intro}

The search for life on other worlds looms large in NASA's 30-year Strategic Plan.\cite{2014arXiv1401.3741K,2015arXiv150704779D} To enable this, NASA is studying a Large UV-Optical-IR Surveyor (LUVOIR) that would use advanced coronagraphs for starlight suppression.\cite{2014arXiv1401.3741K} ATLAST is one concept for LUVOIR. Alternatively, a Habitable Exoplanet Imaging Mission\cite{NASATownhall:2015tm} has been proposed. This might pair a smaller aperture space telescope with a starshade flying in formation. Alternatively, an off-axis non-segmented telescope with a coronagraph has been suggested. In any case, better detectors than exist today are highly desirable.

Our emphasis in this paper is on ATLAST\cite{Bolcar:2015td}, and specifically on ATLAST's detectors for spectroscopic biosignature characterization in the VISIR (hereafter just ``biosignature characterization'') . Although not discussed here, other ATLAST technology needs include precision large-scale optics, ultra-stable structures, starlight suppression, and mirror coatings (See Ref.~\citenum{Bolcar:2015td}). This emphasis sets aside the importance of the UV to ATLAST's overall mission. Within the ATLAST study, detector and other technology development is envisioned across ATLAST's $90~\textrm{nm}-2.5~\mu\textrm{m}$ ``stretch goal'' wavelength range.\cite{Bolcar:2015td}

This paper closely follows the SPIE presentation. We begin with an introduction to biosignature characterization, and show that biosignature characterization is ultra-low background astronomy. The extremely low background count rates motivate the need for further work on VISIR detectors. In Sec.~\ref{ATLASTNeeds}, we briefly summarize ATLAST's VISIR detector needs in the context of existing technologies.

Sec.~\ref{DetStatus} discusses what are arguably the two most mature detector technologies for biosignature characterization in greater detail. These are electron multiplying CCDs (EMCCDs) for the visible and HgCdTe avalanche photodiode (APD) arrays for the VISIR. We also include a more speculative discussion of what might be achieved in conventional HgCdTe arrays with appropriately optimized readout integrated circuits (ROIC). 

\section{BIOSIGNATURE CHARACTERIZATION}\label{BioSig}

\begin{figure}[t]
\begin{center}
\includegraphics[width=0.8\columnwidth]{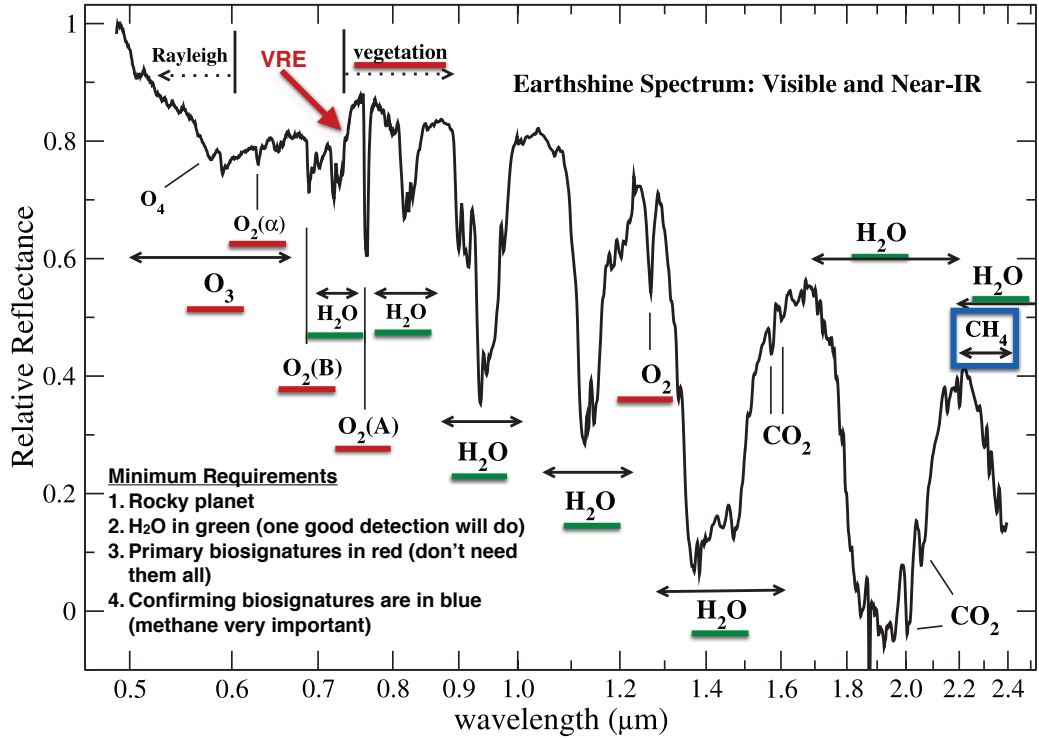}
\caption{This figure overlays a number of important biosignatures on Turnbull~\etal's\cite{2006ApJ...644..551T} spectrum of earth seen as an exoplanet. In addition to the features shown here, there is a strong $\textrm{O}_3$ bandhead at about 260~nm that is considered a primary biosignature. The vegetation red edge (VRE) is caused by chlorophyl from plants. The individual spectral features are discussed in the text.\label{EarthAsExoplanet}}
\end{center}
\end{figure}

Biosignature characterization uses low resolution spectroscopy, $R=\lambda/\Delta\lambda>70$ (required) or $R>150$ (goal), to characterize atmospheric features that are either required for life, or caused by it.\footnote{For purposes of this discussion, the word ``life'' refers to life as we know it on earth today.} Fig.~\ref{EarthAsExoplanet} shows several important biosignatures overlaid on a spectrum of earth as it would appear if seen as an exoplanet.

To make this spectrum, Turnbull~\etal\cite{2006ApJ...644..551T} observed the night side of the moon and used modeling to solve for the earth's contribution as it would appear to a distant observer. We define a life detection as consisting of; (1) a rocky planet, (2) with water vapor, (3) a primary biosignature, and (4) a confirming biosignature to rule out false positives. Lacking a confirming biosignature, one could attempt to increase the statistical significance of a result by resolving the temporal dependence of a feature. Arguments for a biological source could also be strengthened by placing the detection in a more comprehensive geological and astrophysical context by measuring other atmospheric gases including ${\rm CO_2}$, and characterization of the host star's energy distribution.

With regard to false positives, methane is thought to be particularly important because it is difficult to simultaneously maintain significant concentrations of oxygen, ozone, and methane. Non-equilibrium concentrations are most straightforwardly explained by biological processes. The methane feature at 2.32~$\mu$m is unfortunately blended with water vapor. However, there is another methane feature between 3~$\mu$m and 3.5~$\mu$m that might be better if the observatory could observe it. The spectrum shows a few other features, notably ${\rm CO_2}$ and ${\rm O_4}$. Although these features do not provide as much information as the primary and secondary biosignatures, they can still be useful, especially when no confirming biosignature is available.

\section{ULTRA-LOW BACKGROUND}\label{ULB}

Even using a $\geq 8~\textrm{m}$ space telescope, biosignature characterization is extreme ultra-low background astronomy, potentially requiring days to observe each exoEarth candidate. Consider a simple toy model with these assumptions; a perfect coronagraph, 25\% efficient integral field unit (IFU) spectrograph, $\lambda=550$~nm, pixel size $=0.7\times 1.22\lambda/D$, $R=150$, and the background is $3\times$ the earth's zodiacal light. With these assumptions, the background count rate is $<0.001~\textrm{cts}~s^{-1}~\textrm{pix}^{-1}$. More sophisticated models that include the effects of imperfect coronagraphs and simulated exoEarths reach the same conclusion: biosignature characterization is extremely photon starved.\cite{Stark:2015er} 

For such extremely low count rates, a single photon detector (SPD) is clearly preferred. Better than photon counting, an SPD counts individual photons without adding appreciable noise from any source. The needed SPD combines high QE, zero read noise, ultra-low dark current, and ultra-low spurious count rate. In an EMCCD, clock induced charge (CIC) is one example of spurious counts that it would be beneficial to reduce. In IR APD arrays, glow from non-optimized ROICs is another example of spurious counts that it would be beneficial to eliminate. We discuss both EMCCDs and IR APD arrays in more detail later.

\section{ATLAST Detector Needs}\label{ATLASTNeeds}

The ATLAST technology development plan has been discussed elsewhere at this conference.\cite{Bolcar:2015td} Tab.~\ref{ReqTab} summarizes the ATLAST detector needs and ``technology gaps''. Because detectors for the UV through near-IR are equally important to the ATLAST mission, we show them all here. However, this presentation is focused specifically on the VISIR, which we highlight with a red box. The grayed out technologies are no less important to the mission.
\begin{table}[t]
\begin{center}
\caption{Detector Technology Components and Identified ``Gaps''$\rm ^a$}\label{ReqTab}
\includegraphics[width=\columnwidth]{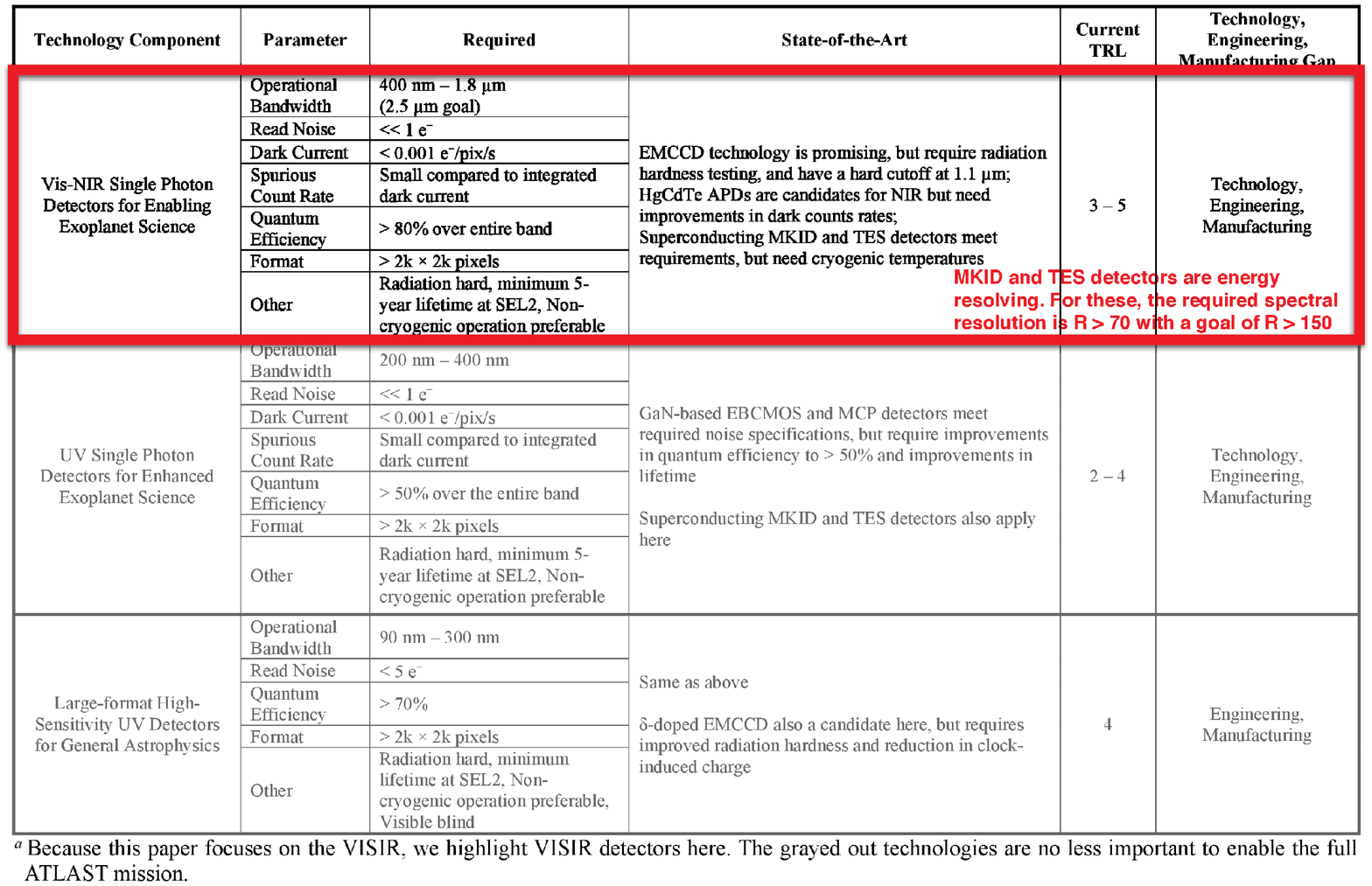}
\end{center}
\end{table}

Broadly speaking, the need is for $\textrm{QE}>80\%$ SPDs from 400~nm through 1.8~$\mu$m (2.5~$\mu$m goal). The $\rm 2K\times 2K$ pixel format is needed if used with an IFU. For space flight, the detectors need to be radiation hard in the L2 radiation environment and able to survive launch loads and vibration. 

The ATLAST study team has indicated a strong preference for non-cryogenic detectors if they can enable the science. This is driven by goals that include: (1) simplifying the system engineering, (2) simplifying the integration and test flow, and (3) completely retiring the risks associated with cooling the detectors to $\textrm{T}\sim 100~\textrm{mK}$.

Coronagraphs capable of achieving contrasts of $10^{-10}$ require wavefront error stability at the level of tens of picometers. Exported vibrations from a conventional cryocooler would present an obvious threat to achieving this. If cryogenic detectors are to be used, then cooling technology development is needed to provide essentially zero vibration cooling. If the cooling challenges could be creatively overcome, then cryogenic detectors including microwave kinetic inductance devices (MKID) and transition edge sensor (TES) arrays might become attractive. 

Once cooled, Both TESs and MKIDs already function as true SPDs with built in energy resolution. Both MKIDs\cite{2013PASP..125.1348M} and TESs\cite{1999ApJ...521L.153R,Romani:2001wl} have been used for refereed astrophysics publications. For biosignature characterization, both would require further development to improve parameters including their VISIR energy resolution and the efficiency of coupling light to detector elements. However, since this publication is about ATLAST, we defer further discussion of cryogenic detectors to a later publication.

Consistent with ATLAST's preference for non-cryogenic detectors if possible, we take it as a requirement that the detectors will be operated at a temperature that can be achieved using only passive cooling, $\textrm{T}\gtrsim 30~\textrm{K}$. In early JWST studies, this emerged as a practical detector temperature that could be achieved with margin at L2.

\section{Candidate VISIR Detector Technologies for ATLAST}\label{DetTech}

Although no completely satisfactory VISIR detector candidate exists for ATLAST today, Tab.~\ref{DetCandidates} summarizes a number of promising technologies. In making this list, we limited consideration to detectors that we believe to be at least NASA TRL-3. This unavoidably leaves some lower TRL, but nevertheless promising technologies off the list. We encourage all efforts that aim to meet the needs outlined in Sec.~\ref{ATLASTNeeds}, even if the specific technology does not appear in this table. Tab.~\ref{DetCandidates} includes a few detectors that we will not be discussing further here because they would operate at $\textrm{T}\leq 30~\textrm{K}$. These are MKIDs, TESs, superconducting nanowire single photon detectors (SNSPD), and Si:As hybrids.
\begin{table}[t]
\begin{center}
\caption{ATLAST VISIR Detector Candidates}\label{DetCandidates}
\includegraphics[width=0.65\columnwidth]{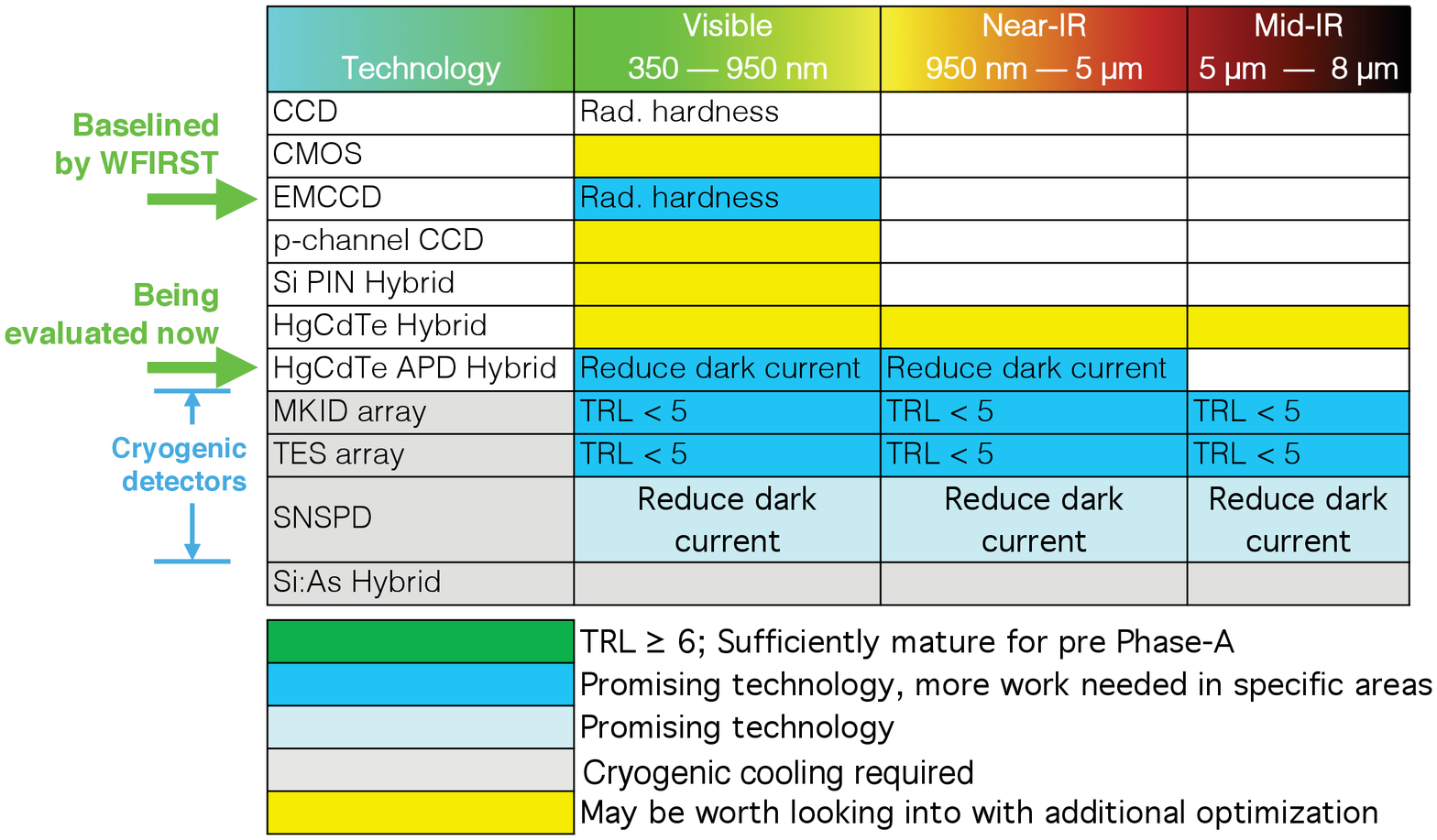}
\end{center}
\end{table}

Tab.~\ref{DetCandidates} attempts to condense a wide trade space into a simple graphic for presentation purposes. Both dark and light blue represent existing technologies that we believe hold significant promise. Dark blue is arguably higher TRL than light blue for biosignature characterization. Yellow indicates a technology that we did not discuss to comply with presentation time limits, but that nevertheless may hold promise for further investigation.

Two of the technologies that we discuss in depth, EMCCDs and IR APDs, are both shaded dark blue. Of these, EMCCDs are currently closer to meeting performance requirements in the visible. IR APDs are the most mature non-cryogenic candidate spanning the VISIR. Although HgCdTe hybrids are shaded yellow in Tab.~\ref{DetCandidates}, we also discuss these in Sec.~\ref{DetStatus} because we plan to investigate them further in our labs at Goddard.

\section{Status of a Few Detector Candidates}\label{DetStatus}

\subsection{EMCCD Status}\label{EMCCDStatus}

For over a decade, EMCCDs have been leading candidates for low background photon counting in the visible. Starting in the early 2000s, several groups have explored individual photon counting with EMCCDs. In 2004, Daigle~\textit{et al.}\cite{Daigle:2004jka} studied how an e2v CCD97 camera, ``operating in pure photon counting mode would behave based on experimental data.'' In 2006, Wen~\textit{et al.}\cite{Wen:2006gv} evaluated an e2v CCD201 for space astronomy and published images of a test pattern showing that the EMCCD operated as a photon counter. Over the ensuing decade, steady progress has been made, and today it is possible to buy a commercial EMCCD camera from NuVu Cameras that uses shaped clocks and high readout rates to achieve $\rm CIC<0.002~cts~pix^{-1}~frame^{-1}$. EMCCDs have been baselined for WFIRST's coronagraph and several presentations at this conference discuss WFIRST's EMCCD efforts. \cite{Harding:2015tz,Bush:2015uf}.

When new and un-degraded by the space radiation environment, EMCCDs are arguably able to meet even ATLAST's challenging performance needs. However, like any conventional n-channel CCD, they will degrade when irradiated. This is a consequence of the phosphorus that is used to dope the n-type channels. Radiation damage, including charge transfer efficiency degradation and pixel operability degradation, has been one of the major reasons for replacing the Hubble Space Telescope's (HST) CCDs. Understanding how radiation effects EMCCDs is important to both WFIRST and ATLAST. Although the ATLAST detector requirements will ultimately be more challenging than those for WFIRST, WFIRST nevertheless provides a valuable early opportunity to understand the issues and address them. For WFIRST, JPL has begun radiation testing and mitigation studies.\cite{Bush:2015uf}

Although EMCCDs are promising detectors for ATLAST, more work in selected areas would be very beneficial. These include efforts aimed at: (1) improving radiation tolerance, (2) further reducing clock induced charge, and (3) improving the red QE from about 850~nm to the bandgap wavelength. Any improvements in radiation tolerance will lead to longer usable life at L2. Further reduction in CIC is important for ATLAST because it is currently a major component of the noise budget. Improving the red QE is important because the strongest water line that is in band for a silicon detector is found at about 950~nm, where the QE of conventional CCDs tends to be falling rapidly.

\subsection{HgCdTe Photodiode Status}\label{MCTStatus}

\begin{figure}[t]
\begin{center}
\includegraphics[width=0.8\columnwidth]{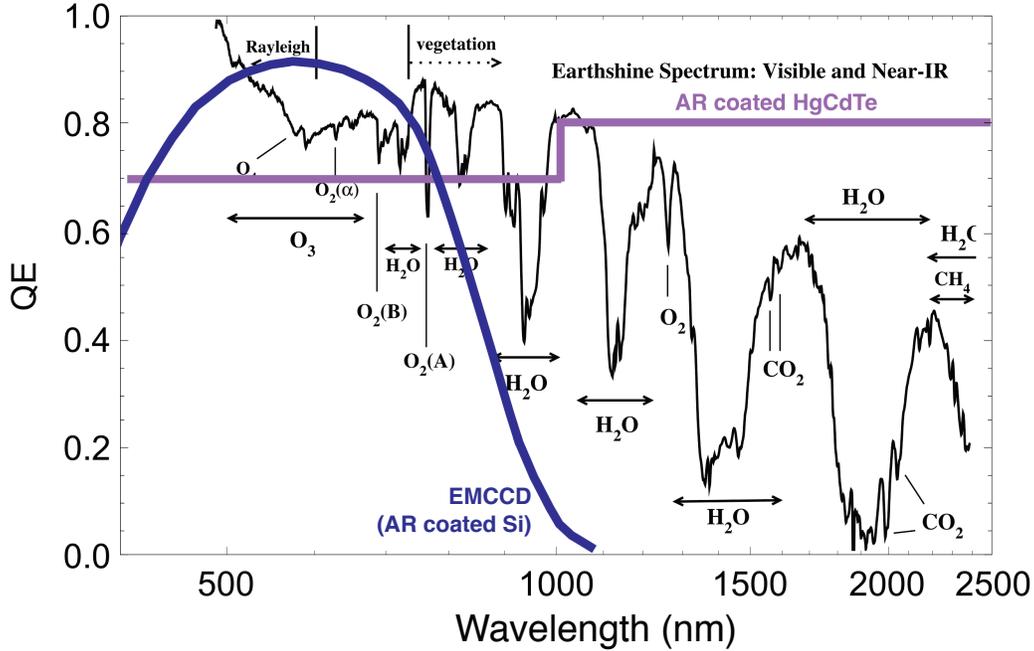}
\caption{AR coated HgCdTe has demonstrated $\rm QE > 70\%$ from $0.6-1~\mu\rm m$ and $\rm QE > 80\%$ from $1-2.5~\mu\rm m$ for JWST. The major vendors claim that using optimized designs, they can extend this performance to about 400~nm. We show the full potential range here, although the QE performance from $400-600$~nm needs to be confirmed in an astronomical detector.\label{QE}}
\end{center}
\end{figure}

Compared to many other detector materials, HgCdTe has shown very good QE from 400~nm through 2.5~$\mu$m and beyond (Fig.~\ref{QE}). For example, the James Webb Space Telescope's (JWST) near-IR detectors achieve $\rm QE>70\%$ from $0.6-1~\mu\rm m$, and $\rm QE>80\%$ from $1-2.5~\mu$m. For non-astronomical applications, the major vendors have delivered HgCdTe detectors that function at wavelengths at least as short as 400~nm. This impressive QE and high overall maturity begs the question, could conventional HgCdTe photodiode arrays achieve much lower noise than they do today if paired with the right readout approach?

\begin{figure}[t]
\begin{center}
\includegraphics[width=0.5\columnwidth]{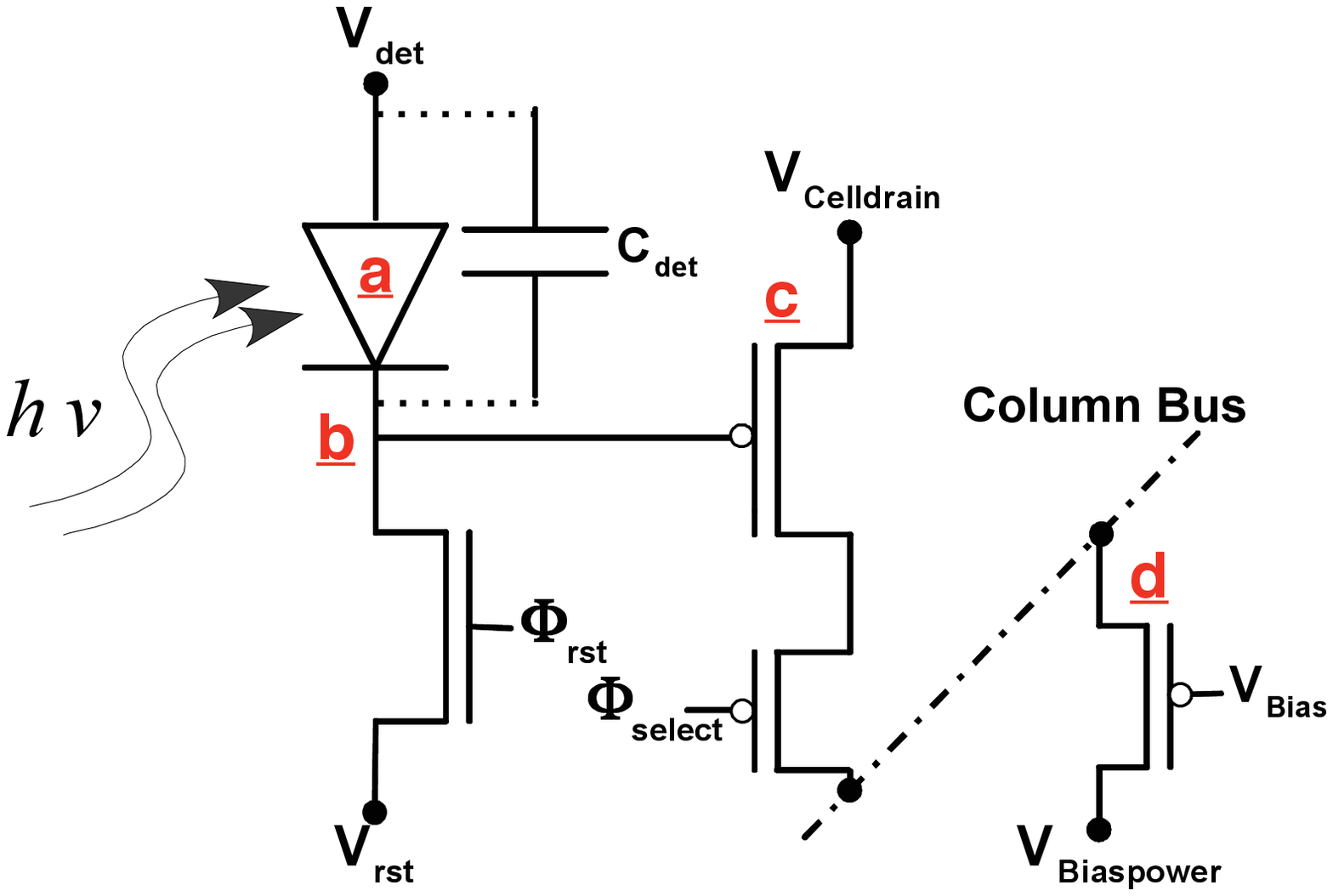}
\caption{Since the mid 1980s, most HgCdTe arrays for low background astronomy have used an SFD architecture to multiplex many pixels onto a few video outputs. However, the overall SFD system may not be optimal for achieving the lowest possible noise. To understand the full potential of HgCdTe photodiode arrays as ultra-low noise detectors, it would be helpful to better understand the noise that originates in: (a) the photodiode itself, (b) the resistive contact, (c) the pixel source-follower, and (d) the output source follower if one is used. Armed with comprehensive understanding of the noise components, it might be possible to design other ROIC architectures to achieve the noise floor that is set by the photodiode alone. This figure is based closely on a corresponding figure from Ref.~\citenum{Loose:2003vh}.\label{SFD}}
\end{center}
\end{figure}

Since the mid-1980s, most low background astronomy arrays have used a source-follower per detector architecture (SFD; Fig.~\ref{SFD}). The SFD architecture has the advantages that it is simple, low power, low glow (when properly designed), and has met performance requirements up to and including those for WFIRST. A major factor driving ROIC design up to the present day has been the need to multiplex a large number of pixels out through a small number of video outputs. This necessitates very wide measurement bandwidth in the video electronics to reproduce the complicated waveform as the output changes from pixel to pixel.

A good start toward understanding the full potential of HgCdTe photodiode arrays as ultra-low noise detectors would be detailed characterization of existing JWST and WFIRST arrays aimed at separating the noise contributions from elements $a-e$ of Fig.~\ref{SFD}. The aim would be an itemized noise budget rather than the lumped ``read noise'' that is conventionally reported.

Although SFD arrays are well adapted to many kinds of astronomy, the current SFD design may not be optimal for achieving the lowest possible noise. The fundamental noise floor of an \eg JWST HgCdTe photodiode (a) is potentially of order $\sqrt{i_d t}$, where $i_d$ is the dark current and $t$ is integration time. The JWST NIRCam arrays have $i_d\sim 0.001~e^-~s^{-1}~\textrm{pixel}^{-1}$. Although conventional HgCdTe photodiode arrays will never function as ideal SPDs on account of leakage current at temperatures $\rm T>30~K$, it is possible that today's H2RG and H4RG detectors are not yet approaching the fundamental noise limits of the photodiodes themselves. It would be interesting to see what could be achieved if noise from the resistive interconnect (b) could be reduced, and/or different ROICs and readout strategies could substantially reduce or eliminate $1/f$ noise from (c) the pixel source follower. The output source follower (d) can already be bypassed in many cases.

\subsection{IR APD Status}\label{IRAPDStatus}

HgCdTe APD arrays are a promising technology that initially entered astronomy for comparatively high background applications including adaptive optics and interferometry\cite{Finger:2012dv} and wavefront sensing and fringe tracking\cite{Finger:2014we}. More recently, they have been used at the telescope to provide diffraction-limited imaging via the ``lucky imaging'' technique.\cite{2014SPIE.9154E..19A} Although HgCdTe APD arrays have been made by DRS, Raytheon, and Teledyne; those made by Selex are the focus of most attention in astronomy now.

A group at the University of Hawaii has been evaluating Selex SAPHIRA arrays for applications that include low background astronomy.\cite{2014SPIE.9154E..19A} With appropriately optimized process, the HgCdTe itself is probably capable of the same QE performance as the JWST arrays. Moreover, because gain is built into the pixels before the first amplifier, they promise photon counting and potentially even single photon detection if ``dark current'' can be reduced to acceptable levels.

``Dark current'' is the most significant obstacle to using Selex APD arrays for ultra-low background astronomy today. The $\sim 10-20~e^-~s^{-1}~\textrm{pixel}^{-1}$ gain corrected ``dark current'' that has been reported\cite{2014SPIE.9154E..19A} is almost certainly dominated by glow from the ROIC. The ROIC in current devices was not optimized for ultra-low background, or even low background astronomy. Work continues at the University of Hawaii to try to disentangle ROIC glow from more fundamental leakage currents in current generation APD arrays. On the longer term, work is also underway aimed at optimizing the ROIC design.

Although HgCdTe APD arrays hold out the promise of read noise below that which can be achieved using conventional photodiode; like conventional photodiodes there will ultimately be a leakage current noise floor that is determined by thermally activated defect states in the HgCdTe. However, it is likely that today's performance is still far from that floor, and more work is needed to better understand the full potential of HgCdTe APD arrays for ultra-low background astronomy in the context of missions like ATLAST.

\section{Summary}\label{Summary}

The search for life on other worlds promises to be a major focus for astrophysics in coming decades. In space, the tools will include new observatories like ATLAST equipped with high performance coronagraphs and/or starshades. These will enable biosignature characterization of exoEarths.

Fortunately, good detector prototypes exist, although more work is needed to mature them for ATLAST. In the VISIR, these include EMCCDs and IR APD arrays. More speculatively, HgCdTe photodiode arrays may still have room for improvement, even beyond the impressive performance that has been shown for JWST and that is expected for WFIRST. Although the challenges are real, they are solvable. To get from where we are today to what is needed to fully enable ATLAST, focused strategic investment in VISIR detectors is needed.

\acknowledgments
This work was supported by NASA as part of the Goddard Space Flight Center Science and Exploration Directorate Life Finder Detectors Science Task Group (STG). 


\end{document}